\documentclass[aps,prd,preprint,groupedaddress,showpacs]{revtex4}

\usepackage{epsfig}
\usepackage{graphics}
\usepackage{slashed}
\usepackage{amssymb,amsmath}
\begin{document}

\leftmargin -2cm
\def\choosen{\atopwithdelims..}

\title{Inclusive $b$-jet and $b\bar b$-dijet production at the LHC \\ via Reggeized gluons}

\author{\firstname{V.A.} \surname{Saleev}}
\email{saleev@samsu.ru}

\affiliation{Samara State University, Academic Pavlov Street 1,
443011 Samara, Russia}

%\author{\firstname{A.V. }\surname{Shipilova}}
%\email{alexshipilova@ssu.samara.ru}
%\affiliation{Samara State University, Ac.~Pavlov, 1, 443011, Samara, Russia}

\author{\firstname{A.V. }\surname{Shipilova}}\email{alexshipilova@samsu.ru}
\affiliation{ Institut f. Kernphysik, Forschungszentrum Juelich,
52425 Juelich, Germany}
 \affiliation{ Samara State University,
Academic Pavlov Street 1, 443011 Samara, Russia}

\begin{abstract}
We study inclusive $b$-jet and $b\bar b$-dijet production at the
CERN LHC invoking the hypothesis of gluon Reggeization in
$t$-channel exchanges at high energy. The $b$-jet cross section
includes contributions from open $b$-quark production and from
$b$-quark production via gluon-to-bottom-pair fragmentation. The
transverse-momentum distributions of inclusive $b$-jet production
measured with the ATLAS detector at the CERN LHC in different
rapidity ranges are calculated both within multi-Regge kinematics
and quasi-multi-Regge kinematics. The $b\bar b$-dijet
cross-section is calculated within quasi-multi-Regge kinematics as
a function of the dijet invariant mass $M_{jj}$, the azimuthal
angle between the two jets $\Delta\phi$ and the angular variable
$\chi$. At the numerical calculation, we adopt the
Kimber-Martin-Ryskin and Bl\"umlein prescriptions to derive
unintegrated gluon distribution function of the proton from its
collinear counterpart, for which we use the
Martin-Roberts-Stirling-Thorne set.  We find good agreement with
measurements by the ATLAS and CMS Collaborations at the LHC at the
hadronic c.m.\ energy of $\sqrt S=7$~TeV.
\end{abstract}

%\date{16.01.2012}
\maketitle

\section{Introduction}
\label{sec:one}

The study of $b$-jet hadroproduction provides an important test of
perturbative quantum chromodynamics (QCD) at high energies. The
total collision energies, $\sqrt{S}=1.8$~TeV and 1.96~TeV in
Tevatron runs~I and II, respectively, and $\sqrt{S}=7$~TeV or
14~TeV at the LHC, sufficiently exceed the characteristic scale
$\mu$ of the relevant hard processes, which is of order of $b$-jet
transverse momentum $p_T$, {\it i.e.}\ we have
$\Lambda_\mathrm{QCD}\ll\mu\ll\sqrt{S}$. In this high-energy
regime, so called "Regge limit", the contribution of partonic
subprocesses involving $t$-channel parton (gluon or quark)
exchanges to the production cross section can become dominant.
Then the transverse momenta of the incoming partons and their
off-shell properties can no longer be neglected, and we deal with
"Reggeized" $t$-channel partons. These $t-$channel exchanges obey
multi-Regge kinematics (MRK), when the particles produced in the
collision are strongly separated in rapidity. If the same
situation is realized with groups of particles, then
quasi-multi-Regge kinematics (QMRK) is at work. In the case of
$b$-jet and $b\bar b$-dijet inclusive production, this means the
following: $b$-jet (MRK) or $b\bar b$-dijet (QMRK) is produced in
the central region of rapidity, while other particles are produced
with large modula of rapidities.

The parton Reggeization approach \cite{QMRK} is based on the
hypothesis of parton Reggeization in $t-$channel exchanges at high
energy \cite{BFKL}. It was used for the description of a large
number of hard processes at the modern hadron colliders and the
obtained results confirm the assumption of a dominant role of MRK
or QMRK production mechanisms at high energy. This approach was
successfully applied to interpret the production of isolated jets
\cite{KSS2011}, prompt photons \cite{SVADISy}, diphotons
\cite{SVAdiy}, charmed mesons \cite{PRD}, heavy quarkonia
\cite{PRD2003,KSVcharm,KSVbottom,Psi2012} measured at the Fermilab
Tevatron, at the DESY HERA and at the CERN LHC. The theoretical
background of a parton Reggeization approach is the effective
quantum field theory implemented with the non-Abelian
gauge-invariant action including fields of Reggeized gluons
\cite{BFKL} and Reggeized quarks \cite{LipatoVyazovsky}, which was
proposed by L.~N.~Lipatov in 1995 \cite{Lipatov95}. In this
effective theory Reggeized partons interact with quarks and
Yang-Mills gluons in a specific way. Recently, in
Ref.~\cite{Antonov}, the Feynman rules for the effective theory of
Reggeized gluons were derived for the induced and some important
effective vertices.

Usually it is suggested that MRK or QMRK production mechanism to
be the dominant one only at small $p_T$ values. Our recent study
of isolated jet production at the Tevatron and LHC colliders, see
Ref.~\cite{KSS2011}, demonstrated that the parton Reggeization
approach can be successfully used already in the range of
$x_T=\frac{2 p_T}{\sqrt{S}} \lesssim 0.1$, or at the $p_T\lesssim
300-400$~GeV for the energy $\sqrt{S}=7$~TeV at the LHC. This
result motivates us to apply the parton Reggeization approach for
the study of $b$-jet and $b\bar b$-dijet production in the
kinematical range of transverse momentum $20<p_T<400$~GeV and
rapidity $|y|<2.1$, as it was measured by the ATLAS Collaboration
at the CERN LHC \cite{ATLASb}.

The high-energy factorization scheme with the effective vertices
for Reggeized gluons has been used  earlier in
Refs.~\cite{Teryaevb,PRb} for description of inclusive open
$b$-quark \cite{CDFBmeson}, $b$-jet \cite{CDFb1} and $b\bar
b$-dijet \cite{CDFb2} production at the Tevatron collider. In this
paper, we study in the same manner the inclusive $b$-jet and
$b\bar b$-dijet production at the CERN LHC invoking the hypothesis
of gluon Reggeization in $t$-channel exchanges at high energy. We
take into account two mechanisms of $b$-jet production: the open
$b$-quark production and "jet-like" $b$-quark production via
gluon-to-bottom-pair fragmentation \cite{FrixioneMangano}. We
consider $b$-quark jet as an isolated, by the jet-cone condition
\cite{Rcone}, hadronic jet containing one $b(\bar b)$-quark or
$b\bar b$-quark pair. Thus, the $b$-jet production cross section
can be written as a sum of two terms. The first one represents a
so-called "open $b$-quark" production, when the $b$-jet contains
$b(\bar b)$-quark which is produced directly in the hard partonic
subprocess. The second term corresponds to the case of "jet-like"
production, where a $b$-jet contains $b\bar b$-quark pair which is
produced via gluon or light-quark fragmentation. The
transverse-momentum distributions of inclusive $b$-jet production
measured with the ATLAS detector at CERN LHC \cite{ATLASb} in the
different rapidity ranges are calculated both within multi-Regge
kinematics and quasi-multi-Regge kinematics. The $b\bar b$-dijet
cross sections are calculated within quasi-multi-Regge kinematics
as functions of the $b\bar b$-dijet invariant mass $M_{jj}$, the
azimuthal angle between the two jets $\Delta\phi$ and the angular
variable $\chi$.

This paper is organized as follows. In Sec.~\ref{sec:two} the
parton Reggeization approach is briefly reviewed. We write down
the relevant for our analysis analytical formulas for squared
matrix elements and differential cross sections. In
Sec.~\ref{sec:three}, we describe our calculations and present the
results obtained. In Sec.~\ref{sec:four}, the conclusions are
summarized.

\section{Model}
\label{sec:two}
 We study $b$-jet production in the region of large
$b$-quark transverse momentum $p_T\gg m_b$, where $m_b$ is a
$b$-quark mass. At the present time, the conventional approach for
calculation of the $b-$quark production cross sections is based on
the next-to-leading (NLO) approximation in perturbative QCD and
collinear parton model \cite{FMNR97}. It is well known that
fixed-order perturbation QCD calculations are applicable when the
transverse momentum $p_T$ of the produced heavy $b-$quark is not
much larger than its mass $m_b$. In the case when the transverse
momentum significantly exceeds the mass, the large logarithms of
type $\log(p_T/m_b)$ arise to all orders of $\alpha_s(\mu)$, so
that a fixed-order approach breaks down~ \cite{NasonDawsonEllis}.
It is possible to resum all these logarithms in the fragmentation
approach using the factorization theorem, which states that the
cross section for the production process of high-$p_T$ $b-$quark
can be written in factorized form as a convolution of the
short-distance partonic cross section of parton $f$ production
with the fragmentation function $D_f^b(z,\mu^2)$ for a formation
of a $b-$quark from the parton $f$:
\begin{equation}
\frac{d\hat\sigma^{frag}}{dp_T}=\sum_f \int dz\int dp~'_T
\frac{d\hat\sigma^f}{dp~'_T}D_{f}^b(z,\mu)\delta(p_T-z
p~'_T).\label{eq:frag1}
\end{equation}
The fragmentation functions for heavy quarks in perturbative QCD
have been studied at the next-to-leading order (NLO) QCD approach
in Ref.~\cite{MeleNason}.

The experimentally measured transverse energy $E_T$ (or the
transverse momentum $p_T$) of $b$-jet includes transverse energies
(transverse momenta) of all partons inside some jet-cone in the
rapidity--azimuthal angle plane, which radius is defined as
follows, $R=\sqrt{\triangle y^2+\triangle\phi^2}$ \cite{Rcone}.
Such a way, it is insignificant which part of the initial parton
four-momentum is transferred to the $b$-quark, and we can simplify
the formula (\ref{eq:frag1}) to the form
\begin{equation}
\frac{d\hat\sigma^{frag}}{dp_T}=\sum_f
\frac{d\hat\sigma^f}{dp_T}n_{f}(\mu),\label{eq:frag2}
\end{equation}
where $n_{f}(\mu)=\int_0^1D_{f}^b(z,\mu)dz$ is a $b$-quark
multiplicity in the $f$-parton jet. It is obvious that a $b$-quark
multiplicity in a gluon-initiated jet greatly exceeds a $b$-quark
multiplicity in any quark-initiated jets, $n_{g}(\mu)\gg
n_{q}(\mu)$ with $q=u,d,s,c$. We will take into account only main
contribution from the gluon-to-bottom-pair fragmentation $g\to
b\bar b$. Let us note that in this case the $b\bar b$-pair is
considered as a one $b$-quark jet.

To describe inclusive $b$-jet and $b\bar b$-jet cross sections in
terms of the parton Reggeization approach, in the LO we need to
consider gluon fusion subprocesses of open $b$-quark and gluon
production only, which are to be dominant at the high energy, they
write:
\begin{eqnarray}
R(q_1)+R(q_2)&\to& g(p),\label{RRg}\\
R(q_1)+R(q_2)&\to& b(p_1)+\bar b(p_2),
\label{RRbb}\\
R(q_1)+R(q_2)&\to& g(p_1)+g(p_2),\label{RRgg}
\end{eqnarray}
where $R$ is a Reggeized gluon and $g$ is a Yang-Mills gluon,
respectively, with four-momenta indicated in parentheses. The
contribution of the partonic subprocess (\ref{RRgg}) can be
neglected in comparison with the contribution of the subprocess
(\ref{RRbb}) because of the strong suppression by the $g\to b\bar b$
fragmentation ($n_g\thickapprox 10^{-3}$) for both produced gluons.
In Ref.~\cite{PRb} it was shown, that at the Tevatron energy range,
the contribution of the subprocesses $Q+\bar Q\to g$ and $Q+\bar
Q\to b+\bar b$ with initial Reggeized quarks is sufficiently smaller
comparing to the dominant contribution of the subprocesses
(\ref{RRg}) and (\ref{RRbb}), and the former becomes sizeable only
at the very large $b$-jet transverse momentum $p_T$. As the LHC
energy exceeds by a factor $3.5$ the one of the Tevatron collider,
we estimate a quark-antiquark annihilation contribution to be even
much smaller and therefore do not consider it in the present
analysis.
%%%%%%%%%%%%%%%%%%%%%%%%%%%%%%%%%%%%%%%%%%%%%%%%%%%%%%%%%%
%
%    First correction
%
%%%%%%%%%%%%%%%%%%%%%%%%%%%%%%%%%%%%%%%%%%%%%%%%%%%%%%%%%%%

{ Performing a study of high-transverse-momentum $b-$quark
production ($p_T>>m_b$) in the collinear parton model, we have an
additional $b-$quark production mechanism, namely a production via
$b-$flavor excitation, where $b(\bar b)-$quarks are considered as
partons in the colliding protons. For example, this mechanism has
been used successfully to describe $B-$meson $p_T-$spectra at the
Tevatron and LHC in NLO calculations of the parton model
\cite{Kniehl_Bmeson}. We have used a similar idea in our previous
study of inclusive $b-$jet production at the Tevatron within the
Parton Reggeization Approach \cite{PRb}. In this work we took into
account the LO in $\alpha_s$ contribution from $2\to 1$ partonic
subprocess
\begin{equation}
{\cal B}R\to b,\label{bRb}
\end{equation}
where ${\cal B}$ is a Reggeized $b-$quark. As it is shown in the
Fig.~1 of Ref.~\cite{PRb}, the sum of this contribution and a
contribution from the subprocess (\ref{RRbb}) strongly
overestimates the experimental data.  In the present analysis we
ignore a contribution from the subprocess (\ref{bRb}). First, we
avoid any chance of a double-counting between subprocesses
(\ref{amp:RRbb}) and (\ref{bRb}). Second, the conception of quark
Reggeization for a $b-$quark inside a proton seems to be wrong.
$b-$quarks are produced preferably at the last step of
QCD-evolution at the large scale $\mu\sim p_T$, and their PDF is
proportional to a large logarithm $\log (p_T/m_b)$. However, the
QCD-evolution of a Reggeized parton should be valence-like. It
means, the Reggeized parton must be a $t-$channel parton
throughout all steps of QCD-evolution in the parton ladder. But, a
$b-$quark conventional collinear PDF, which we take as input for a
KMR (Bl\"umlein) prescription to obtain a $b-$quark unintegrated
PDF, satisfies sea-like QCD-evolution. For this reason we strongly
overestimate a value of a $b-$quark unintegrated PDF. The more
adequate way should be to consider a subprocess $bR\to b$ with a
collinear $b-$quark in the initial state, instead of subprocess
(\ref{bRb}). But even in this case a problem of double-counting
still exist. That is why in the present study we consider
Reggeized-gluon induced contributions, like (\ref{RRg}) and
(\ref{RRbb}), only. }
%%%%%%%%%%%%%%%%%%%%%%%%%%%%%%%%%%%%%%%%%%%%%%

The squared amplitude of subprocess~(\ref{RRg}) reads
\cite{PRD2003,PRb}:
\begin{eqnarray}
\overline{|{\cal M}(R+R\to g)|^2}=\frac{3}{2}\pi \alpha_s {p}_T^2,
\label{equ:ampRQ}
\end{eqnarray}
where  $p_T^2=t_1+t_2+2\sqrt{t_1t_2}\cos \phi_{12}$,
$t_1=-q_1^2={\vec q}_{1T}^{~2}$, $t_2=-q_2^2={\vec q}_{2T}^{~2}$,
 with ${\vec q}_{1T}$ and ${\vec q}_{2T}$
representing the transverse momenta of initial Reggeized gluons,
and $\phi_{12}$ is the azimuthal angle enclosed between them.

The squared amplitude of subprocess (\ref{RRbb}) was obtained in
Ref.~\cite{SaleevVasinBc} using the effective Feynman rules of the
parton Reggeization approach. It coincides with previous result of
Ref.~\cite{Collins} which is expressed in the alternative form.
The answer of Refs.~\cite{SaleevVasinBc,Collins} can be written
down as a linear combination of an Abelian and a non-Abelian term,
as
\begin{equation}
\overline{|{\cal M}(R + R \to b + \bar b)|^2} = 256 \pi^2
\alpha_s^2 \left[ \frac{1}{2 N_c} {\cal M}_{\mathrm{A}} +
\frac{N_c}{2 (N_c^2 - 1)} {\cal M}_{\mathrm{NA}} \right],
\label{amp:RRbb}
\end{equation}
where
\begin{eqnarray}
{\cal M}_{\mathrm{A}}&=&\frac{t_1 t_2}{{\tilde t} {\tilde u}} -
\left( 1 + \frac{\alpha_1\beta_2 S}{\tilde
u}+\frac{\alpha_2\beta_1S}{\tilde t} \right)^2,
\nonumber\\
{\cal M}_{\mathrm{NA}} &=&
\frac{2}{S^2}\left(\frac{\alpha_1\beta_2S^2}{{\tilde
u}}+\frac{S}{2}+\frac{\Delta}{\hat
s}\right)\left(\frac{\alpha_2\beta_1S^2}{{\tilde
t}}+\frac{S}{2}-\frac{\Delta}{\hat s}\right)\nonumber\\
&&{}-\frac{t_1 t_2}{x_1x_2{\hat s}}\left[\left(\frac{1}{{\tilde
t}}-\frac{1}{{\tilde
u}}\right)(\alpha_1\beta_2-\alpha_2\beta_1)+\frac{x_1x_2 {\hat
s}}{{\tilde t} {\tilde u}}-\frac{2}{S}\right],
\nonumber\\
\Delta&=&\frac{S}{2}\left[{\tilde u} - {\tilde t}+2
S(\alpha_1\beta_2-\alpha_2\beta_1) +t_1
\frac{\beta_1-\beta_2}{\beta_1+\beta_2} -t_2
\frac{\alpha_1-\alpha_2}{\alpha_1+\alpha_2}\right],
\label{amp:RRbbDelta}
\end{eqnarray}
$\tilde t = \hat t - m_b^2$, $\tilde u = \hat u - m_b^2 $, $\alpha_1=2(p_1\cdot P_2)/S$,
$\alpha_2=2(p_2\cdot P_2)/S$, $\beta_1=2(p_1\cdot P_1)/S$, and
$\beta_2=2(p_2\cdot P_1)/S$. Here, the Mandelstam variables are
defined as $\hat s = (q_1 + q_2)^2$, $\hat t = (q_1 - p_1)^2$, $\hat
u = (q_2 - p_1)^2$, $S=(P_1+P_2)^2$, where $P_1$ and $P_2$ denote
the four-momenta of the incoming protons.

Exploiting the hypothesis of high-energy factorization, we express
the hadronic cross sections $d\sigma$ as convolutions of partonic
cross sections $d\hat\sigma$ with unintegrated PDFs $\Phi_g^h$ of
Reggeized gluon in the hadrons $h$. For the processes under
consideration here, we have
\begin{eqnarray}
d\sigma(pp\to g X)&=&\int\frac{dx_1}{x_1}\int
\frac{d^2q_{1T}}{\pi}\int\frac{dx_2}{x_2}\int \frac{d^2q_{2T}}{\pi}
\nonumber\\
&&{}\times \Phi^p_{g}(x_1,t_1,\mu^2)\Phi^{ p}_{g}(x_2,t_2,\mu^2)
d\hat\sigma(RR\to g),\label{section1}
\\
d\sigma(pp\to b\bar b X)&=&\int\frac{dx_1}{x_1}\int
\frac{d^2q_{1T}}{\pi}\int\frac{dx_2}{x_2}\int
\frac{d^2q_{2T}}{\pi}
\nonumber\\
&&{}\times \Phi^p_{g}(x_1,t_1,\mu^2)\Phi^{ p}_{g}(x_2,t_2,\mu^2)
d\hat\sigma(RR\to b\bar b)\label{section2}.
\end{eqnarray}
The unintegrated PDFs $\Phi_g^h(x,t,\mu^2)$ are related to their
collinear counterparts $F_g^h(x,\mu^2)$ by the normalization
condition
\begin{equation}
xF_a^h(x,\mu^2)=\int^{\mu^2}dt\,\Phi_a^h(x,t,\mu^2),
\end{equation}
which yields the correct transition from formulas in the parton
Reggeization approach to those in the collinear parton model,
where the transverse momenta of the partons are neglected.

In our numerical analysis, we adopt as our default the prescription
proposed by Kimber, Martin, and Ryskin (KMR) \cite{KMR} to obtain
unintegrated gluon PDF of the proton from the conventional
integrated one, as implemented in Watt's code \cite{Watt}. { The
precise analysis of KMR gluon unintegrated PDF had been performed in
the Ref. \cite{KMRstudy}, including an accurate study of the
dependence on the choice of collinear input.} As is well known
\cite{Andersson:2002cf}, other popular prescriptions, such as those
by Bl\"umlein \cite{Blumlein:1995eu} or by Jung and Salam
\cite{Jung:2000hk}, produce unintegrated PDFs with distinctly
different $t$ dependences.
%%%%%%%%%%%%%%%%%%%%%%%%%%%%%%%%%%%%%%
{ In our analysis we don't evaluate  the unintegrated gluon PDF
after Jung and Salam \cite{Jung:2000hk} because this PDF had been
tabulated only in a range of $t,\mu^2\leq 10^4$ GeV$^2$. It is not
enough to calculate $b-jet$ production cross sections up to
$p_T=400$ GeV, in accordance with measurements of the relevant
experiments. In fact, we had to use the unintegrated gluon PDF up to
$t,\mu^2\leq 10^6$ GeV$^2$. }
%%%%%%%%%%%%%%%%%%%%%%%%%%%%%%%%%%%%%%
In order to assess the resulting theoretical uncertainty, we also
evaluate the unintegrated gluon PDF using the Bl\"umlein approach,
which resums small-$x$ effects according to the
Balitsky-Fadin-Kuraev-Lipatov (BFKL) equation \cite{BFKL}. As input
for these procedures, we use the LO set of the
Martin-Roberts-Stirling-Thorne (MRST) \cite{MRST2002} proton PDF as
our default.
%%%%%%%%%%%%%%%%%%%%%%%%%%%%%%%%%%%%%%
{ The relevant theoretical study of particle production in the
high-energy factorization scheme using KMR and Bl\"umlein
unintegrated gluon PDFs \cite{SVADISy}-\cite{KSVbottom} demonstrates
that both unintegrated PDFs lead to a similar behavior of production
spectra at the non-large particle transverse momentum ($p_T\leq 20$
GeV). In case of high transverse momentum production of isolated
jets and prompt photons \cite{KSS2011}, the theoretical predictions
obtained with these PDFs are different. Although we take identical
collinear inputs for both KMR and Bl\"umlein approaches, the
relevant kernels of integrand transformation between collinear and
unintegrated PDFs differ. The KMR approach is based on DGLAP
evolution equation, while the Bl\"umlein approach is based on BFKL
evolution equation. As the BFKL approach seems to be preferable in
the region of very small $x\ll 1$, which corresponds to non-large
$p_T$ at fixed $\sqrt{S}$, the KMR unintegrated gluon PDF should be
more suitable to describe the experimental data at large $p_T$. }
%%%%%%%%%%%%%%%%%%%%%%%%%%%%%%%%%%%%%%

Throughout our analysis the renormalization and factorization scales
are identified and chosen to be $\mu=\xi p_T$, where $\xi$ is varied
between 1/2 and 2 about its default value 1 to estimate the
theoretical uncertainty due to the freedom in the choice of scales.
The resulting errors are indicated as shaded bands in the figures.

The master formula for the doubly differential cross section of
inclusive $b-$jet production via gluon-to-bottom-pair fragmentation
at the $p_T\gg m_b$ reads as follows
\begin{eqnarray}
\frac{d\sigma^{frag}(pp\to b X)}{dp_Tdy}&=&\frac{1}{p_T^3}\int d
\phi_1\int dt_1
\Phi^p_{g}(x_1,t_1,\mu^2)\Phi^{ p}_{g}(x_2,t_2,\mu^2)\times\nonumber\\
&&\times n_{g}(\mu) \overline{|{\cal M}(RR\to g)|^2} ,
\end{eqnarray}
where $y$ is the rapidity of $b-$quark, $\phi_1$ is the azimuthal angle enclosed
between the vectors $\vec q_{1T}$ and $\vec p_T$,
$$x_{1,2}=\frac{p_T\exp(\pm y)}{\sqrt{S}},
\quad t_2=t_1+p_T^2-2\sqrt{t_1} p_T\cos (\phi_1).$$

In case of $b\bar b-$dijet production via the partonic subprocess
(\ref{RRbb}) we get the differential cross section in the form:
\begin{eqnarray}
\frac{d\sigma^{open}(pp\to b\bar
bX)}{dp_{1T}dy_1dp_{2T}dy_2d\triangle\phi}=\frac{p_{1T}p_{2T}}{16\pi^3}\int
dt_1\int d\phi_1 \Phi^p_{g}(x_1,t_1,\mu^2)\Phi^{
p}_{g}(x_2,t_2,\mu^2) \frac{\overline{|{\cal M}(RR\to b\bar
b)|^2}}{(x_1x_2S)^2},
\end{eqnarray}
where $p_{1,2T}$ and $y_{1,2}$ are $b-$quark and $\bar b-$antiquark
transverse momenta and rapidities, respectively,  $\triangle\phi$ is
the azimuthal angle enclosed between the vectors $\vec p_{1T}$ and
$\vec p_{2T}$, $$x_1=(p_1^0+p_2^0+p_1^z+p_2^z)/\sqrt{S}, \quad
x_2=(p_1^0+p_2^0-p_1^z-p_2^z)/\sqrt{S},$$
$$p_{1,2}^0=\frac{p_{1,2T}}{2}\left[\exp(y_{1,2})+\exp(-y_{1,2})\right],
\quad p_{1,2}^z=\frac{p_{1,2T}}{2}[\exp(y_{1,2})-\exp(-y_{1,2})].$$

The inclusive $b$-jet transverse-momentum spectrum can be presented
in the form:
\begin{equation}
\frac{d\sigma^{bjet}}{dp_T}=\frac{d\sigma^{frag}}{dp_T}+2\times
\frac{d\sigma^{open}}{dp_T}\theta(R_{b\bar b}-R)+
\frac{d\sigma^{open}}{dp_T}\theta(R-R_{b\bar b}),\label{sumbjet}
\end{equation}
where $R_{b\bar b}=\sqrt{(y_b-y_{\bar b})^2+(\phi_b-\phi_{\bar
b})^2}$, $R$ is the experimentally fixed jet radius parameter,
$\theta(x)$ is the unit step function. In such a way, the
subprocess (\ref{RRbb}) of open $b$-quark production contributes
two separate $b$-quark jets while $R_{b\bar b}>R$, and only one
$b$-quark jet while $R_{b\bar b}<R$.

\section{Results}
\label{sec:three}

Recently, the ATLAS Collaboration presented data on inclusive and
dijet production cross sections which have been measured for jets
containing $b$-hadrons ($b$-jets) in proton-proton collisions at a
center-of-mass energy of $\sqrt{S}=7$ TeV \cite{ATLASb}. The
inclusive $b$-jet cross section was measured as a function of
transverse momentum in the range $20<p_T<400$ GeV and rapidity in
the range $|y|<2.1$. The $b\bar b$-dijet cross section was measured
as a function of the dijet invariant mass in the range
$110<M_{jj}<760$ GeV, the azimuthal angle difference between the two
jets $\Delta\phi$ and the angular variable $\chi$ in two dijet mass
regions. Jets were reconstructed with jet radius parameter $R=0.4$.
The angular variable $\chi$ is defined as follows $\chi=\exp
|y_1-y_2|$. To measure the cross section as a function of $\chi$, an
additional acceptance requirement was used that restricts the boost
of the dijet system to $|y_{boost}|=0.5 |y_1+y_2|<1.1$.

%%%%%%%%%%%%%%%%%%%%%%%%%%%%%%%%%%%%%%%%%%%%%%%%%%%%%%%%%%

\begin{figure}[ht]
\begin{center}
\includegraphics[width=.8\textwidth, clip=]{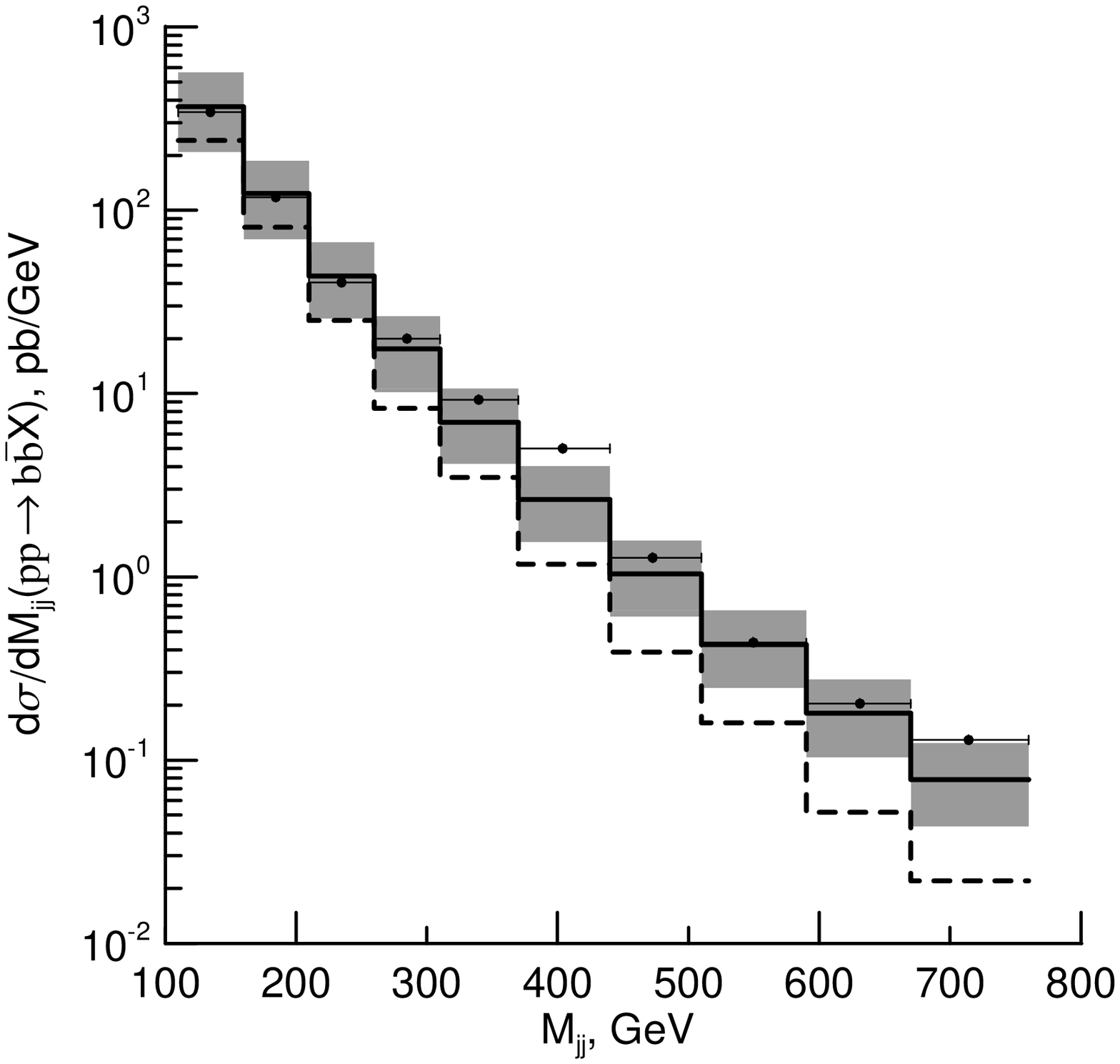}
\end{center}
\caption{\label{fig:1} The $b\bar b$-dijet cross-section as a
function of dijet invariant mass $M_{jj}$ for $b$-jets with
$p_T>40$~GeV, $|y|<2.1$. The data are from ATLAS Collaboration
\cite{ATLASb}, the solid polyline corresponds to KMR unintegrated
PDF, the dashed one
--- to Bl\"umlein PDF. The shaded bands indicate the theoretical
uncertainties in the case of KMR unintegrated PDF.}
\end{figure}

The $b\bar b$-dijet cross-section as a function of dijet invariant
mass $M_{jj}$ for $b$-jets with $p_T>40$~GeV and $|y|<2.1$ is shown
in Fig.~\ref{fig:1}. The data are compared to LO parton Reggeization
approach predictions, the solid polyline corresponds to KMR
unintegrated PDF \cite{KMR}, the dashed one --- to Bl\"umlein PDF
\cite{Blumlein:1995eu}. We observe nice agreement between data and
theoretical prediction obtained with the KMR unintegrated PDF. In
case of Bl\"umlein PDF, the theoretical histogram lies about factor
2 lower than the experimental data and this difference increases
towards the high values of dijet invariant mass.

\begin{figure}[ht]
\begin{center}
\includegraphics[width=.8\textwidth, clip=]{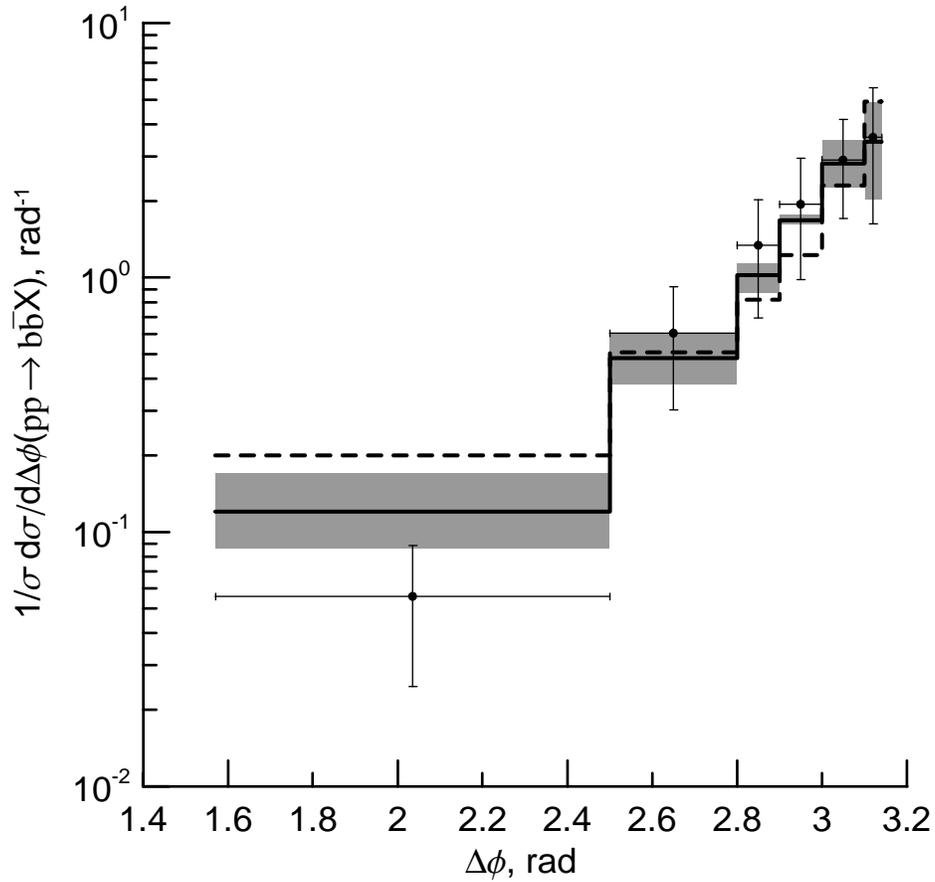}
\end{center}
\caption{\label{fig:2} The $b\bar b$-dijet cross-section as a
function of the azimuthal angle difference between the two jets for
$b$-jets with $p_T>40$~GeV, $|y|<2.1$ and a dijet invariant mass of
$M_{jj}<110$~GeV. The data are from ATLAS Collaboration
\cite{ATLASb}, the solid polyline corresponds to KMR unintegrated
PDF, the dashed one --- to Bl\"umlein PDF. The shaded bands indicate
the theoretical uncertainties in the case of KMR unintegrated PDF.}
\end{figure}

In Fig.~\ref{fig:2} the $b\bar b$-dijet cross-section as a function
of the azimuthal angle difference $\Delta \phi$ between the two jets
for $b$-jets with $p_T>40$~GeV, $|y|<2.1$ and a dijet invariant mass
of $M_{jj}>110$~GeV is presented. The normalized to the total cross
section data are compared to LO parton Reggeization approach
predictions, the solid polyline corresponds to KMR unintegrated PDF,
the dashed --- to Bl\"umlein PDF. For the both unintegrated PDFs our
predictions lie within the experimental uncertainty interval of data
except only one point at the $\Delta\phi\approx 2$. We need to
mention that in the case of CDF measurements at the Tevatron
\cite{CDFb2} the azimuthal-separation-angle distribution of
inclusive $b\bar b$-dijet production is well described using the
parton Reggeization approach formalism at the all values of the
azimuthal angle difference $0<\Delta\phi<\pi$ (see Fig.~4 in the
Ref.~\cite{PRb}).

\begin{figure}[ht]
\begin{center}
\includegraphics[width=.8\textwidth, clip=]{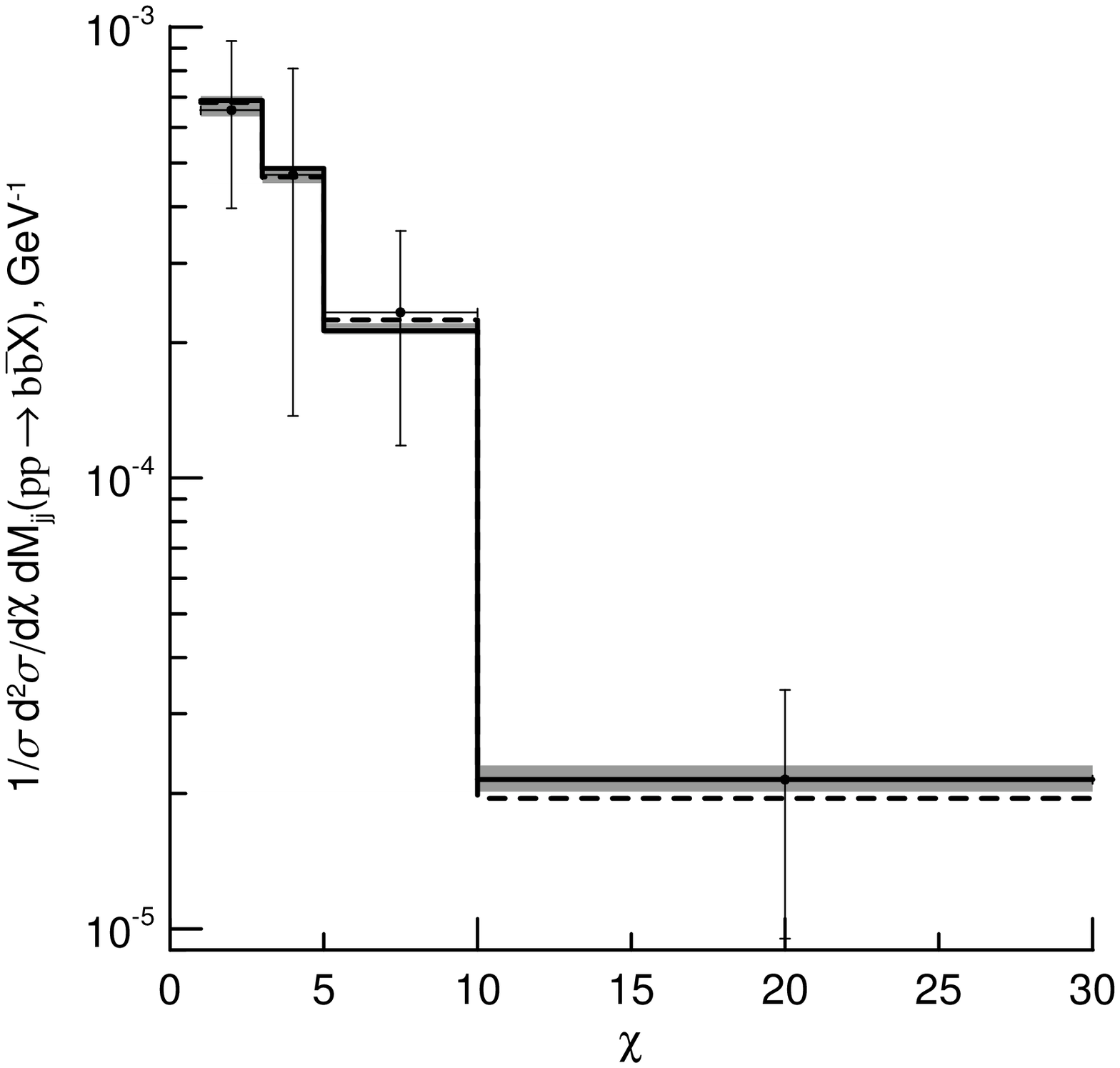}
\end{center}
\caption{\label{fig:3} The $b\bar b$-dijet cross-section as a
function of $\chi$ for $b$-jets with $p_T>40$~GeV, $|y|<2.1$ and
$|y_{boost}|=\frac{1}2|y_1+y_2|<1.1$, for dijet invariant mass range
$110<M_{jj}<370$~GeV. The data are from ATLAS Collaboration
\cite{ATLASb}, the solid polyline corresponds to KMR unintegrated
PDF, the dashed one --- to Bl\"umlein PDF. The shaded bands indicate
the theoretical uncertainties in the case of KMR unintegrated PDF.}
\end{figure}

\begin{figure}[ht]
\begin{center}
\includegraphics[width=.8\textwidth, clip=]{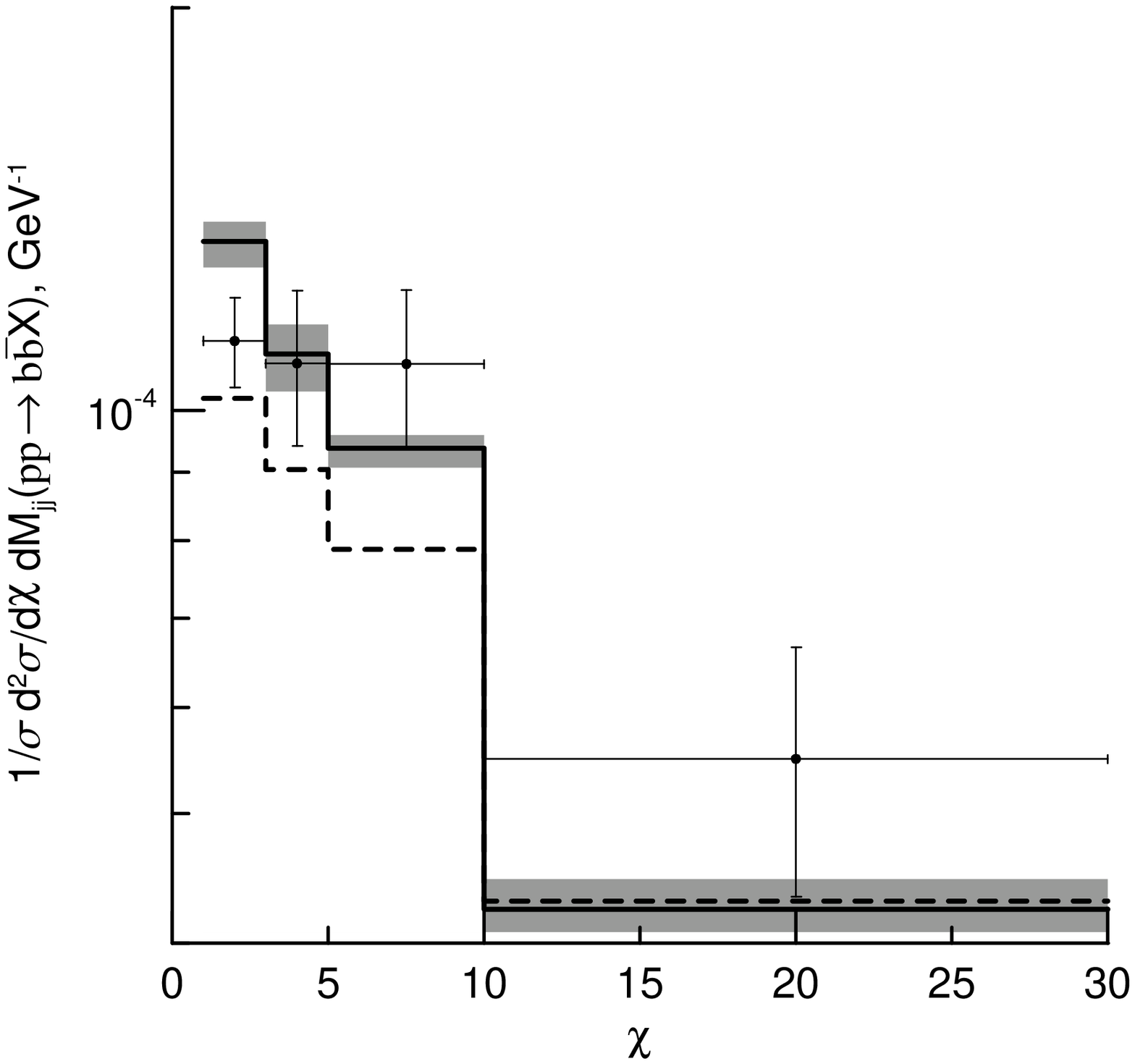}
\end{center}
\caption{\label{fig:4} The $b\bar b$-dijet cross-section as a
function of $\chi$ for $b$-jets with $p_T>40$~GeV, $|y|<2.1$ and
$|y_{boost}|=\frac{1}2|y_1+y_2|<1.1$, for dijet invariant mass range
$370<M_{jj}<850$~GeV. The data are from ATLAS Collaboration
\cite{ATLASb}, the solid polyline corresponds to KMR unintegrated
PDF, the dashed one --- to Bl\"umlein PDF. The shaded bands indicate
the theoretical uncertainties in the case of KMR unintegrated PDF.}
\end{figure}

The $b\bar b$-dijet cross-section as a function of angular
variable $\chi$ for $b$-jets with $p_T>40$~GeV, $|y|<2.1$ and
$|y_{boost}|=\frac{1}2|y_1+y_2|<1.1$, for dijet invariant mass
ranges $110<M_{jj}<370$~GeV and $370<M_{jj}<850$~GeV are shown in
the Fig.~\ref{fig:3} and Fig.~\ref{fig:4}, correspondingly. The
normalized to the total cross section data are compared to our LO
parton Reggeization approach predictions. In the range of
$110<M_{jj}<370$ GeV the polylines corresponding to KMR and to
Bl\"umlein unintegrated PDFs coincide. In the region of large
invariant masses $370<M_{jj}<850$~GeV, the prediction obtained
with the Bl\"umlein unintegrated PDF lies about factor 2 lower
than the data. On the contrary, the calculations with the KMR
unintegrated gluon PDF are found to be in a good agreement with
data.
%%%%%%%%%%%%%%%%%%%

To calculate inclusive $b$-jet transverse-momentum production
spectra we need to take into account gluon-to-bottom-pair
production mechanism and  to use the function of $b\bar b$-pair
multiplicity $n_g(\mu)$ in a gluon jet. Because the existing
theoretical predictions (see, for example,
Ref.~\cite{ManganoNason}) contain large uncertainties, we consider
$n_g(\mu)$ as a free phenomenological parameter, which is
extracted from the experimental data from the ATLAS Collaboration
\cite{ATLASb} for the inclusive $b$-jet cross sections.

\begin{figure}[ht]
\begin{center}
\includegraphics[width=.8\textwidth, clip=]{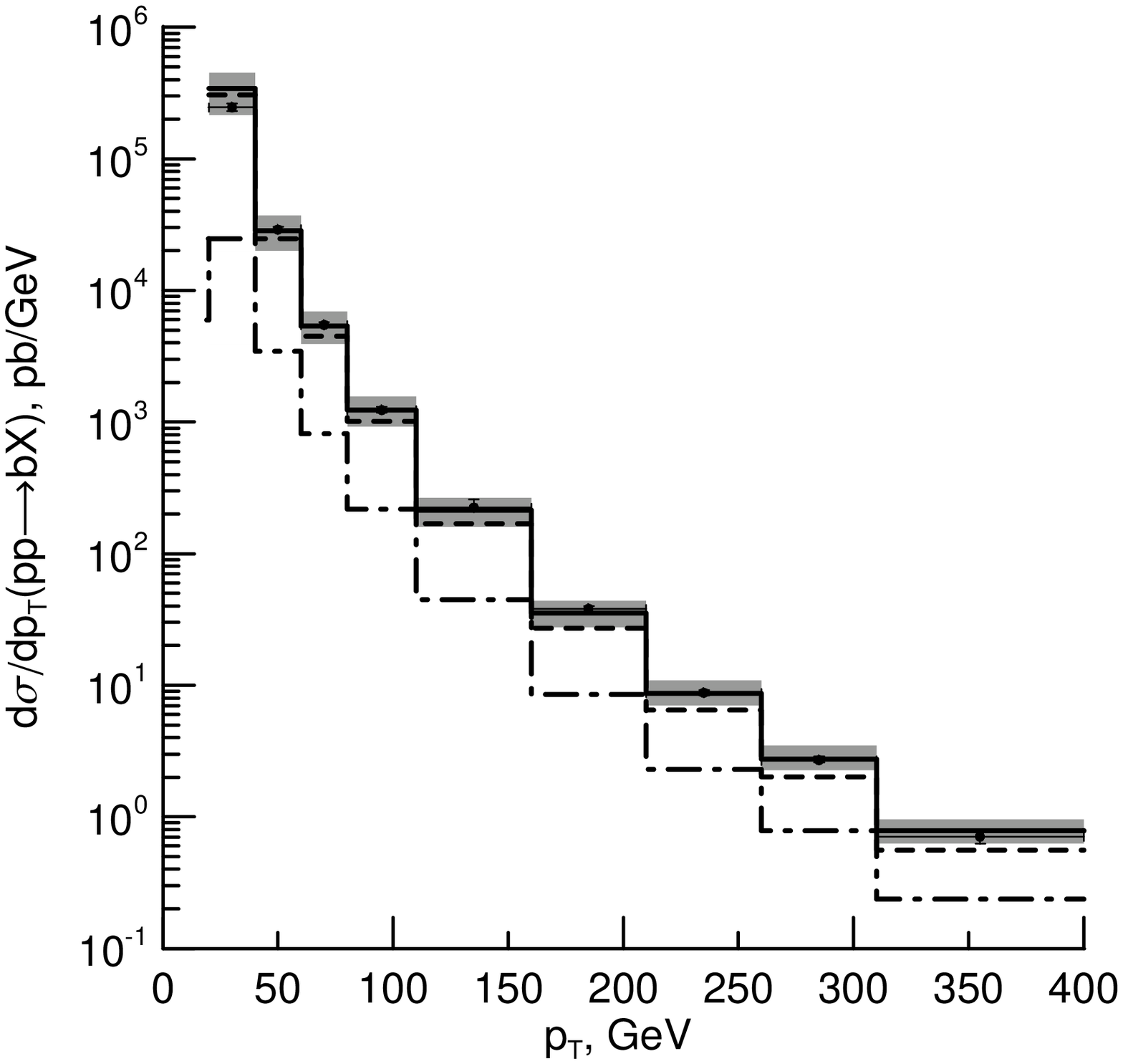}
\end{center}
\caption{\label{fig:5}%
Inclusive differential $b$-jet cross-section as a function of
$p_T$ for $b$-jets with $|y|<2.1$. The data are from ATLAS
Collaboration \cite{ATLASb}. The dashed  polyline corresponds to
contribution of the open $b$-quark production, the dashed-dotted
one --- the gluon-to-bottom-pair fragmentation, the solid ---  sum
of their all. The calculation is done with the KMR unintegrated
PDF.}
\end{figure}

\begin{figure}[ht]
\begin{center}
\includegraphics[width=.8\textwidth, clip=]{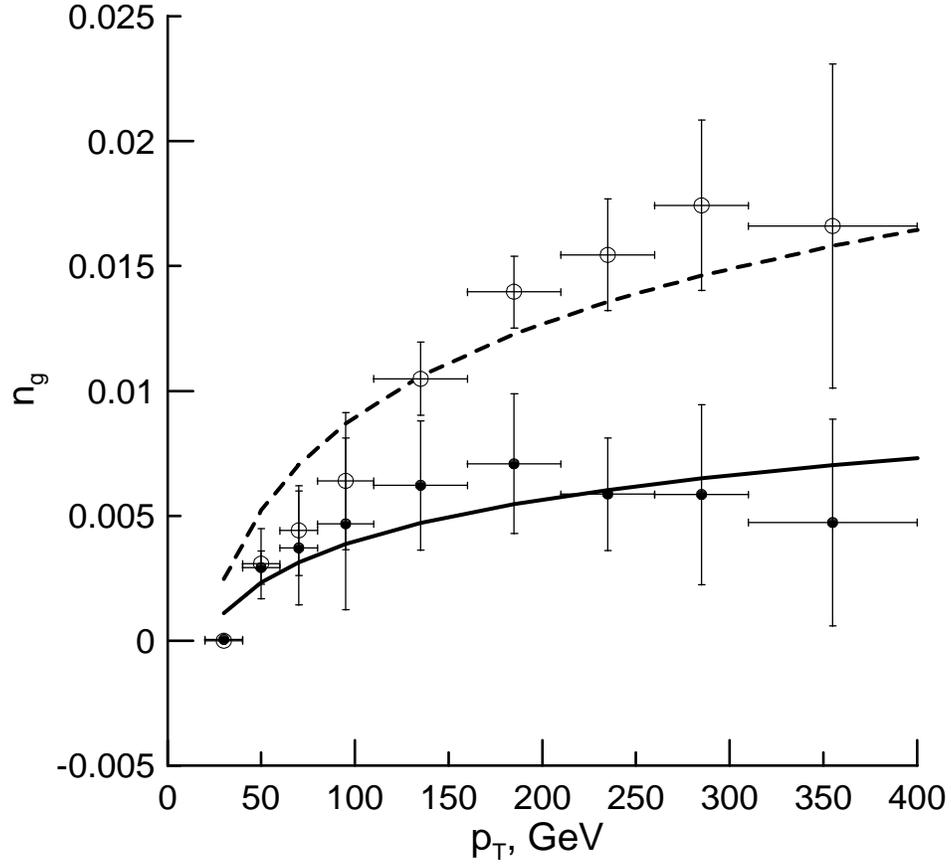}
\end{center}
\caption{\label{fig:6}%
The $b\bar b$-pair multiplicity $n_g$ in a gluon jet  as a function
of $p_T$ extracted from the ATLAS data for the inclusive $b$-jet
production spectra \cite{ATLASb}. The open circles and dashed
fitting line correspond to Bl\"umlein unintegrated PDF, the black
circles and solid fitting line correspond to KMR unintegrated PDF. }
\end{figure}

In the Fig.~\ref{fig:5}, the inclusive differential $b$-jet cross
section as a function of $p_T$ for $b$-jets with $|y|<2.1$ is
compared with our LO predictions of the parton Reggeization
approach. The contribution of QMRK subprocess (\ref{RRbb}) and the
contribution of MRK subprocess (\ref{RRg}) are shown separately.
We see that open $b$-quark production mechanism does not describe
data, especially at the large $p_T$ and some contribution from
gluon-to-bottom-pair fragmentation mechanism is needed. We have
obtained good description of the data using $n_g(\mu)$ as a free
parameter. In Fig.~\ref{fig:6}, the $b\bar b$-pair multiplicity
$n_g(\mu)$ in a gluon jet as a function of $p_T$ extracted from
the ATLAS data for the inclusive $b$-jet production spectra
\cite{ATLASb} is shown. The open circles and dashed fitting line
correspond to Bl\"umlein unintegrated PDF, the black circles and
solid fitting line correspond to KMR unintegrated PDF. The general
theoretical consideration \cite{ManganoNason} leads to the
following analytical approximation for the $b\bar b$-pair
multiplicity in a gluon jet
\begin{equation}
n_g(\mu)= A \ln \frac{\mu^2}{m_b^2},\label{ngfit}
\end{equation}
where we fixed $m_b=4.75$ GeV and $\mu=p_T/4$ \cite{Matteo2004},
and found that $A_{KMR}=0.0012$ in case of KMR unintegrated PDF,
and $A_{B}=0.0027$ in case of Bl\"umlein unintegrated PDF. At the
scale $\mu\backsimeq m_Z/4$, which corresponds
gluon-to-bottom-pair fragmentation of secondary gluon in the
$Z$-boson decay ($Z\to q\bar q\to q\bar q b\bar b$), our
approximation (\ref{ngfit}) yields $n_g\simeq 0.002-0.004$, that
is in an agreement with the measurements at the LEP Collider:
$n_g=(3.3\pm 1.8)\times 10^{-3}$ from the DELPHI Collaboration~
\cite{LEP1}, $n_g=(2.44\pm 0.93)\times 10^{-3}$ from the SLD
Collaboration~\cite{LEP2}. The difference in obtained $b\bar
b$-pair multiplicities $n_g(\mu)$ with the KMR and Bl\"umlein
unintegrated PDFs should be used to distinguish last ones. We
conclude that KMR unintegrated PDF looks preferably to describe
$b$-jet production cross sections. Opposite this conclusion, we
found recently \cite{KSS2011} that Bl\"umlein unintegrated PDF is
better to describe inclusive all-flavor inclusive jet production
spectra \cite{Atlas}.

\begin{figure}[ht]
\begin{center}
\includegraphics[width=0.8\textwidth, clip=]{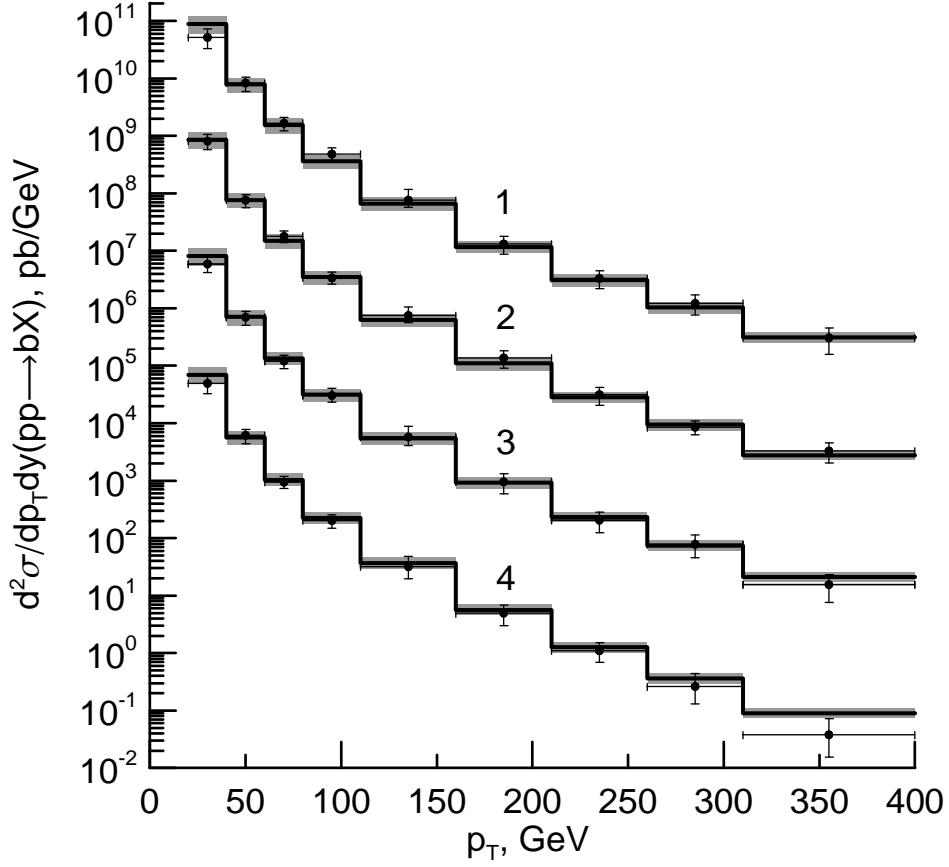}
\end{center}
\caption{\label{fig:7}%
Inclusive double-differential $b$-jet cross-sections as a functions
of $p_T$ for the different rapidity ranges: (1) $|y|<0.3$ (${}\times
10^{6}$), (2) $0.3<|y|<0.8$ (${}\times 10^{4}$), (3) $0.8<|y|<1.2$
(${}\times 10^{2}$) and (4) $1.2<|y|<2.1$. The data are from ATLAS
Collaboration \cite{ATLASb}. The solid polylines correspond to sum
of all contributions (\ref{sumbjet}) and KMR unintegrated PDF.}
\end{figure}

The measured by  ATLAS Collaboration \cite{ATLASb} inclusive
double-differential $b$-jet cross-sections as functions of $p_T$ for
the different rapidity ranges: (1) $|y|<0.3$ (${}\times 10^{6}$),
(2) $0.3<|y|<0.8$ (${}\times 10^{4}$), (3) $0.8<|y|<1.2$ (${}\times
10^{2}$) and (4) $1.2<|y|<2.1$ are shown in Fig.~\ref{fig:7}. Here,
our theoretical predictions are obtained taking into account both
contributions, open $b$-quark production and fragmentation
production with $n_g(\mu)$ as in (\ref{ngfit}), and with KMR
unintegrated PDF. We demonstrate good agreement with data in all
rapidity intervals.

\begin{figure}[ht]
\begin{center}
\includegraphics[width=.9\textwidth, clip=]{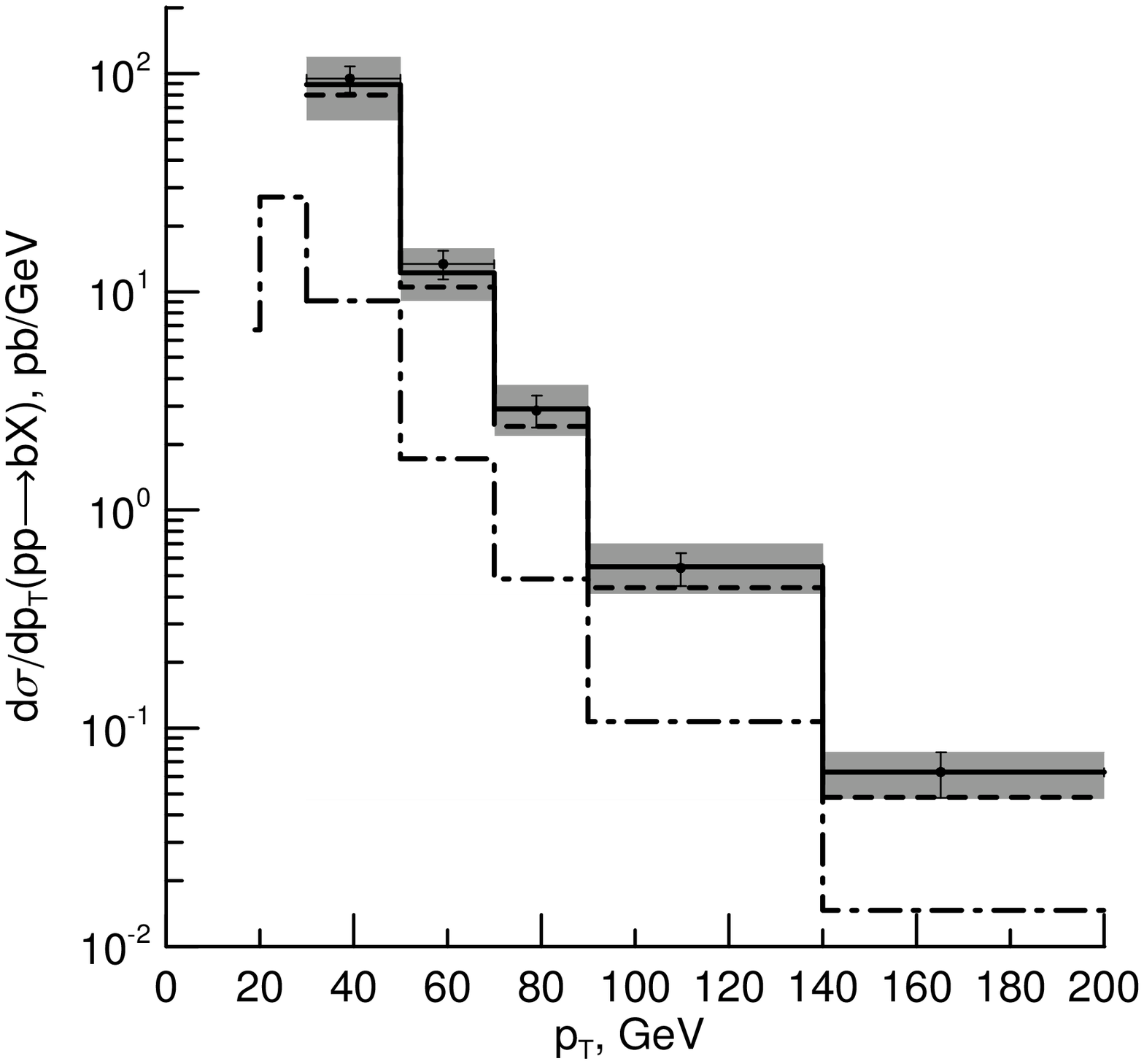}
\end{center}
\caption{\label{fig:8}%
Inclusive differential $b$-jet cross-section as a function of
$p_T$ for $b$-jets with $|y|<2.4$. The data are from CMS
Collaboration \cite{CMSb}. The dashed polyline corresponds to
contribution of the open $b$-quark production, the dashed-dotted
one --- the gluon-to-bottom-pair fragmentation, the solid ---  sum
of their all. The calculation is done with the KMR unintegrated
PDF.}
\end{figure}

\begin{figure}[ht]
\begin{center}
\includegraphics[width=.8\textwidth, clip=]{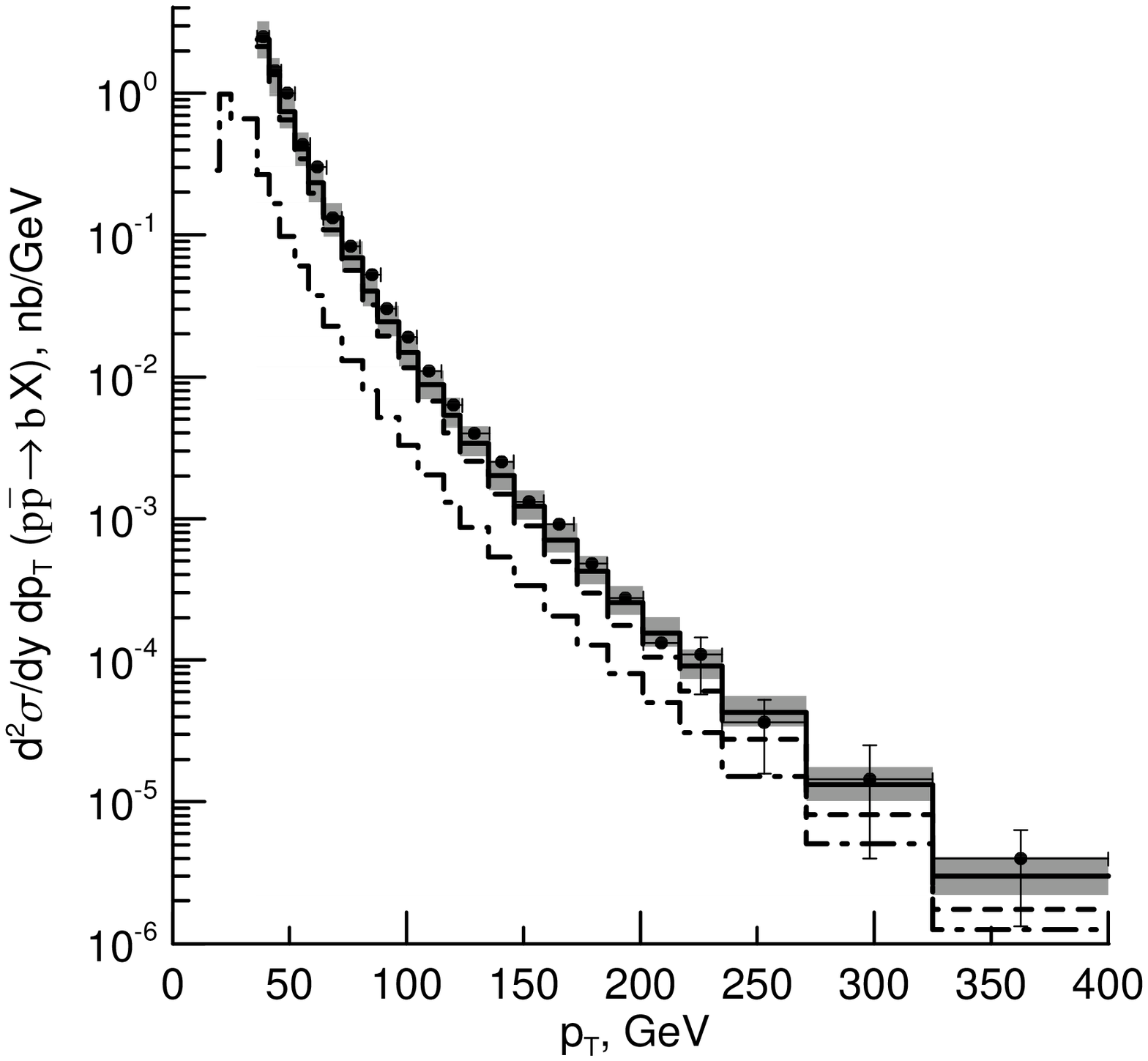}
\end{center}
\caption{\label{fig:9}%
Inclusive differential $b$-jet cross-section as a function of
$p_T$ for $b$-jets with $|y|<0.7$. The data are from CDF
Collaboration \cite{CDFb1}. The dashed polyline corresponds to
contribution of the open $b$-quark production, the dashed-dotted
one --- the gluon-to-bottom-pair fragmentation, the solid ---  sum
of their all. The calculation is done with the KMR unintegrated
PDF.}
\end{figure}

To test the universality of the approach as well as the
universality of the obtained function $n_g(\mu)$ we compare our
prediction with experimental data for transverse-momentum $b$-jet
spectra from  CMS Collaboration at the CERN LHC~\cite{CMSb}
(Fig.~\ref{fig:8}) and CDF Collaboration  at the Fermilab
Tevatron~\cite{CDFb1} (Fig.~\ref{fig:9}). In both cases we find a
good agreement between theoretical predictions and experimental
data.

Looking in Figs.~\ref{fig:5} - \ref{fig:9}, we  find that
contribution of the gluon-to-bottom-pair fragmentation in inclusive
$b$-jet production $p_T$-spectra increases from the 10-15 \% at the
$p_T\simeq 50$ GeV up to 30-40 \% at the $p_T\simeq 350$ GeV. This
conclusion contradicts to the prediction of the NLO calculations in
the collinear parton model in which the gluon-to-bottom-pair
fragmentation mechanism would be dominant at the large-$p_T$ region
at the LHC Collider and it would be about 50 \% at the Tevatron
Collider \cite{Matteo2004,JHEP07}.

{ Comparing as a whole our results with theoretical predictions
obtained in the NLO of a parton model, which also describe ATLAS
data for $b$-jet production \cite{ATLASb}, we would like to pay
attention to difficulties of the fixed order collinear calculations.
At first, the K-factor between LO and NLO calculations is very large
at the high $p_T$. The scale uncertainty decreases from LO
calculation to NLO calculation, but it still remains large. The last
one can be a signal on large NNLO contributions, which are not taken
into account. Second, to describe data at non-large $p_T\leq 50$ GeV
at the energy of $\sqrt{S}=2-7$ TeV in the collinear parton model it
is needed to add the soft gluon resummation procedure, which is far
from the application field of DGLAP evolution equation, and should
be considered as a phenomenological trick rather than rigorous
approach. The both these difficulties are solved in the PRA by
introducing the off-shell LO Reggeized parton amplitudes and
unintegrated gluon PDF, which take into account large logarithmic
contributions in all orders in $\alpha_s$: $(\alpha_s\ln
(\mu^2/\Lambda_{QCD}^2))^n$ and $(\alpha_s \ln (1/x))^n$.}

 In such a way, we have obtained the self-coordinated
description in the PRA of $b\bar b$-dijet cross sections, where the
open $b\bar b$-quark pair production works solely, and the $b$-jet
inclusive cross sections, where open $b\bar b$-quark pair production
is the main contribution, while the gluon-to-bottom-pair
fragmentation production is also important.

\section{Conclusions}
\label{sec:four}

The CERN LHC is currently probing particle physics at the
terascale c.m.\ energies $\sqrt{S}$, so that the hierarchy
$\Lambda_\mathrm{QCD}\ll\mu\ll\sqrt{S}$, which defines the MRK and
QMRK regimes, is satisfied for processes of heavy quark ($c$ or
$b$) production in the central region of rapidity, where $\mu$ is
of order of their transverse momentum. In this paper, we studied
QCD processes of particular interest, namely inclusive $b$-jet and
$b\bar b$-dijet hadroproduction, at LOs in the parton Reggeization
approach, in which they are mediated by $2\to1$ and $2\to 2$
partonic subprocesses initiated by Reggeized gluon collisions.

 We describe well recent LHC
data measured by the ATLAS Collaboration~\cite{ATLASb} at the whole
presented range of the $b\bar b$-jet transverse momenta, the $b\bar
b$-jet rapidity, the $b\bar b$-dijet invariant mass $M_{jj}$, the
azimuthal angle between the two jets $\Delta\phi$ and the angular
variable $\chi$. We show that the gluon-to-bottom-pair fragmentation
component \cite{MeleNason}, which takes into account effects of
large logarithms $\log(p_T/m_b)$, increases in inclusive $b-$jet
production at the high transverse momenta $p_T$ up to 30-40 \% of
sum of all contributions. The extracted by the fit of data the
$b\bar b$-pair multiplicity is in agreement with the previous
measurements at the LEP Collider \cite{LEP1,LEP2}.  Comparing
different unintegrated gluon PDFs, we have found that the agreement
with the data has been obtained when we used KMR PDF \cite{KMR}, and
the calculations with Bl\"umlein PDF \cite{Blumlein:1995eu}
regularly underestimate data, approximately by factor 2, in the
region of large $b-$jet $p_T$ and the large $b\bar b-$dijet
invariant mass.

\section{Acknowledgements}

We are grateful to B.~A.~Kniehl, M.~B\"{u}scher and N.~N.~Nikolaev
for useful discussions. The work of A.~V.~S. was supported by the
Federal Ministry for Science and Education of the Russian
Federation under Contract No.~14.740.11.0894 and in part by the
Grant DFG 436 RUS 113/940; the work of V.~A.~S. was supported in
part by the Russian Foundation for Basic Research under Grant
11-02-00769-a.

\newpage

\end{document}